# Thermodynamic Phase Diagrams in λ-$(BETS)_2GaCl_{4-x}Br_x$ under Physical and Chemical Pressure

**Aulia Anisa Firdaus[1*], Arifa Diana Agustin[1,3], Fathan Muyassar Santana[1], Taiyo Oka[2], Shuhei Fukuoka[2], Atsushi Kawamoto[2], Dita Puspita Sari[1,3,4] and Yasuyuki Ishii[1]**

[1]Graduate School of Science and Engineering, Shibaura Institute of Technology, Saitama, 337-8570, Japan
[2] Graduate School of Science, Hokkaido University, 060-0810 Hokkaido, Japan
[3] Nuclear Structure Research Group, RIKEN Nishina Center, 351-0198 Saitama, Japan
[4]Innovative Global Program, College of Engineering, Shibaura Institute of Technology, 337-8570 Saitama, Japan

*E-mail: na24101@shibaura-it.ac.jp

**Abstract.** The evolution of electron correlation in the superconducting state of the mixed-crystal series λ-$(BETS)_2GaBr_xCl_{4-x}$ was investigated by magnetic susceptibility and transport measurements. Br concentrations, independently evaluated by electron-probe microanalysis (EPMA), agree well with values from single-crystal X-ray diffraction refinement. Demagnetization-corrected susceptibility and field-dependent magnetization were measured for single crystals with $x_{c(rf)}$=0.22, 0.24, 0.42, and 0.63, all of which exhibit a diamagnetic superconducting transition. The lower critical field, $H_{c1}$, was determined from the deviation of the magnetization from the Meissner response using the Bean critical-state model, after demagnetization correction; the upper critical field, $H_{c2}$, was determined from the midpoint of the resistive transition in newly measured high-field magnetoresistance data — the first direct measurement of $H_{c2}$ in this mixed-crystal system. $H_{c1}$ shows a dome-*like* maximum at $x_{c(rf)}$=0.24, whereas $H_{c2}$ decreases with increasing Br concentration up to $x_{c(rf)}$=0.42, establishing the high-field superconducting boundary on the chemical-pressure side. Combining these magnetic results with reported transport data under chemical and physical pressure, we propose an electronic pressure–temperature (*P*-*T*) phase diagram for the λ-Ga system. The resulting double-dome-like structure has a local minimum near $x$=0, in the pressure region where a quantum-critical point has previously been proposed at $P$≈0.13 GPa, separating a low-pressure regime where superconductivity coexists with magnetism and an unconventional metallic state from a higher-pressure regime of superconductivity with conventional metallic behavior. This is consistent with the low-temperature resistivity upturn— previously identified only in λ-$(STF)_2GaCl_4$ and λ-$(BETS)_2GaCl_4$ under physical pressure— also being found under chemical pressure, suggesting that the spin-density-wave fluctuations may extend continuously from the chemical- to the physical-pressure axis. This phase diagram overlaps with that of λ-$(STF)_2GaCl_4$ under physical pressure, indicating that chemical substitution, external pressure, and molecular substitution can be integrated into a single bandwidth-controlled phase diagram for λ-type molecular superconductors.

## 1. Introduction

The electronic properties of strongly correlated conductors are generally governed by the electron correlation strength, often expressed as the ratio between the on-site Coulomb repulsion and the electronic bandwidth, $U/W$, and by the band filling, $\delta$ [1]. In high-$T_c$ cuprate superconductors, carrier doping modifies both the band filling and the effective correlation strength [2]. Consequently, changes in carrier concentration and electron correlation are often intertwined, causing the system to evolve through several competing phases, including antiferromagnetic, pseudogap, strange-metal, and superconducting states [3]. Because changes in $\delta$ and $U/W$ occur simultaneously, it is difficult to isolate the effects of electron correlation on the emergence of superconductivity.

Meanwhile, organic superconductors provide a unique platform in which the electronic bandwidth can be tuned while the band filling remains nearly constant, offering a cleaner system for investigating electron correlation, Mott physics, magnetism, and unconventional superconductivity [1]. The κ-$(ET)_2X$ family, where ET denotes bis(ethylenedithio)-tetrathiafulvalene and X is a monovalent anion, is a representative bandwidth-controlled system and has been extensively studied to clarify the interplay between superconductivity and the Mott transition. These materials are effectively half-filled, and their bandwidth can be continuously tuned by external pressure or chemical substitution. Accordingly, bandwidth control can induce a first-order transition between superconducting and Mott-insulating states [4]. A variety of intriguing phenomena have been reported in the κ-$(ET)_2X$ family, including unconventional superconducting states, antiferromagnetic ordering, and spin-liquid behavior [5-9]. These properties are strongly influenced by the anisotropic triangular lattice formed by the κ-type molecular arrangement, in which electron correlations and geometrical frustration play essential roles.

On the other hand, another organic superconductor known as λ-$(BETS)_2GaCl_4$, (BETS: bis(ethylenedithio)tetraselenafulvalene), hereafter abbreviated as λ-Ga, exhibits weaker dimerization than the κ-type molecular arrangement and possesses a lattice structure that combines an anisotropic triangular lattice with a weakly square lattice, offering a unique platform for investigating bandwidth-controlled electron correlations [12]. Furthermore, the superconductivity of λ-Ga was observed at ambient pressure with the superconducting transition temperature ($T_c$) about 6K [10]. Moreover, the unconventional nature of its superconducting state has been suggested both theoretically and experimentally by the deviation of the nodal line structure from conventional *d*-wave symmetry [11-12]. Recent resistivity studies have shown that the superconducting transition temperature, $T_c$, increases with increasing physical pressure, reaching a maximum around $P$=0.13 GPa [13,14]. In addition, the coexistence of superconducting and magnetic phases observed between 0 and 0.13 GPa suggests possible quantum-criticality within this narrow pressure region [14]. However, the comprehensive investigation of the evolution of electron correlation in the superconducting state of λ-Ga under pressure has not yet been reported. Previous phase diagrams of the λ-Ga system in which was discussed within the λ-$(D)_2GaCl_4$ family framework ($D$=ET, STF, BETS) was constructed using the data of mixed-crystal series λ-$(BETS)_2GaBr_xCl_{4-x}$ [15-17]. Additionally, the chemical-pressure range reported in the previous study [15] was not finely spaced compared to our recent resistivity study [18]. In particular, the present resistivity result revealed a broad upturn anomaly at low temperature for low Br concentrations at $x$=0.2. The origin of this anomaly remains unclear. Therefore, the magnetic measurements can provide important insight into whether the upturn anomaly is associated with an enhancement of the non-superconducting fraction at low Br concentration or with other changes in the superconducting properties of the mixed-crystal system.

In this study, we investigate the superconducting properties of λ-$(BETS)_2GaBr_xCl_{4-x}$ by magnetic susceptibility measurement to clarify the evolution of superconductivity under chemical pressure. By combining the present magnetic results with transport data obtained under chemical and physical pressure [14,18], we establish the magnetic-field–pressure ($H$-$P$) phase diagram and propose an electronic pressure–temperature ($P$-$T$) phase diagram for the λ-Ga system. These phase diagrams enable us to examine how chemical and physical pressure influence electron correlations and the stability of the superconducting state in λ-Ga.

## 2. Experimental details

*2.1 Bromine concentration determination using EPMA*

Electron-probe microanalysis (EPMA) was performed using a Shimadzu EPMA-8050G equipped with wavelength-dispersive spectrometers (WDS) to independently determine the Br concentrations in λ-$(BETS)_2GaBr_xCl_{4-x}$. In our previous SEM–EDX study, the S-Kα/Cl-Kα and Se-Lα/Br-Lα signals overlapped, while the Br-Kα line was not clearly observed [19]. Consequently, the total halogen content, $\sum Br, Cl$, obtained for the mixed crystals deviated from the expected stoichiometry value by four [19]. However, SEM–EDX yielded approximately four of Cl atoms for pristine λ-$(BETS)_2GaCl_4$, suggesting that the discrepancy in the total halogen content of the mixed crystals may arise from uncertainty in the Br quantification. Therefore, the higher spectral resolution of WDS enables reliable separation of overlapping elemental peaks, allowing more accurate quantitative analysis of the Br and Cl concentrations.

The quantitative analysis using EPMA was performed on the same single crystal used for resistivity [18], SCXRD [18,20], SEM-EDX [19], and magnetic measurement reported in the present study to ensure the consistency result. The spectra of C-Kα, S-Kα, Cl-Kα, Ga-Kα, Se-Lα, and Br-Lα lines were measured using LSA120, PET, LiF, ADP, and RAP analyzing crystals. The Se-Kα (0.1105 nm) and Br-Kα (0.1040 nm) lines lie below the lower wavelength limit of the LiF crystal (0.12206 nm); therefore, the Se-Lα and Br-Lα lines were used, and their peak-center wavelengths were carefully determined from the WDS peak scans to be 0.89880 and 0.83694 nm, respectively. This enabled them to be distinguished from the neighboring Se-Lβ emission near 0.874 nm, thereby minimizing spectral interference. Similarly, the S-Kα and Cl-Kα lines were well resolved in the WDS peak scans, with experimentally determined peak-center wavelengths of approximately 0.53747 and 0.47295 nm, respectively. Pure C, $Sb_2S_3$, pure Se, GaAs, NaCl, and KBr were used as standards for C, S, Se, Ga, Cl, and Br, respectively. The quantitative measurements were performed at an accelerating voltage of 15 kV and a beam current of 4.1 nA using the minimum beam condition to reduce damage to the organic crystals. In addition, elemental mapping was also performed for the $x_{c(rf)}$ = 0.24 and 0.63 samples. The measured intensities were corrected using the ZAF2 and the resulting atomic percentages were normalized to Ga and denoted as $\mu_N$. The Br occupancy was then calculated using Eq. (2).

$$\mu_N(\mathrm{i}) = \frac{\mu_{(i)}}{\mu_{Ga}} \tag{1}$$

with $\mu_{(i)}$ is the average atomic percentage from each element

$$\text{Br fraction} = \frac{\mu_{Br}}{\mu_{Cl}+\mu_{Br}} = \frac{\mu_{N(Br)}}{\mu_{N(Cl)}+\mu_{N(Br)}}$$

$$\text{Br occupancy} = 4\ x\ Br\ fraction \tag{2a}$$

$$\text{Cl occupancy} = 4\ -\ Br\ occupancy \tag{2b}$$

Here, $\mu_{N(Cl)}$ and $\mu_{N(Br)}$ are the average atomic percentage of Cl and Br element normalized to the Ga atom. The factor of 4 arises to accounts for the Br occupancy across the four halogen sites of the tetrahedral anion $GaBr_xCl_{4-x}$.

*2.2 superconducting magnetic measurements*

Magnetic measurements were performed using an MPMS-XL SQUID magnetometer (Quantum Design). The mixed crystal sample with needle-*like* shape was obtained as a collaboration work with Hokkaido University. The single crystals measured in the present study were selected from the same single crystal as those used for the resistivity measurements reported previously [18], to ensure consistency between the magnetic and transport data. The single crystals measured for SQUID measurement were labelled as $x$= 0.2 S1, $x$ = 0.2 S2, $x$ = 0.4 S1, and $x$ = 0.65. In the previous study [18,19], the composition $x_c$ was evaluated using X-ray diffraction (XRD) measurements. The re-calculation of $x_c$ with refinement of each atomic position within the unit cell was obtained from a separate structural refinement analysis [20]. This refined analysis gives $x_{c(rf)}$= 0.22, 0.24, 0.42, and 0.63 for $x$ = 0.2 S1, $x$ = 0.2 S2, $x$ = 0.4 S1, and $x$ = 0.65, respectively in which will be used throughout the present study when discussing the evolution of superconducting properties in the mixed crystal compound. The single crystals size was 1.44 x 0.18 x 0.05 $mm^3$; 1.14 x 0.11 x 0.05 $mm^3$; 0.54 x 0.16 x 0.09 $mm^3$ and 1.10 x 0.14 x 0.05 $mm^3$, respectively. The length, width, and thickness correspond to the finite length of a-, b, and c- axis from the Figure 1(c).

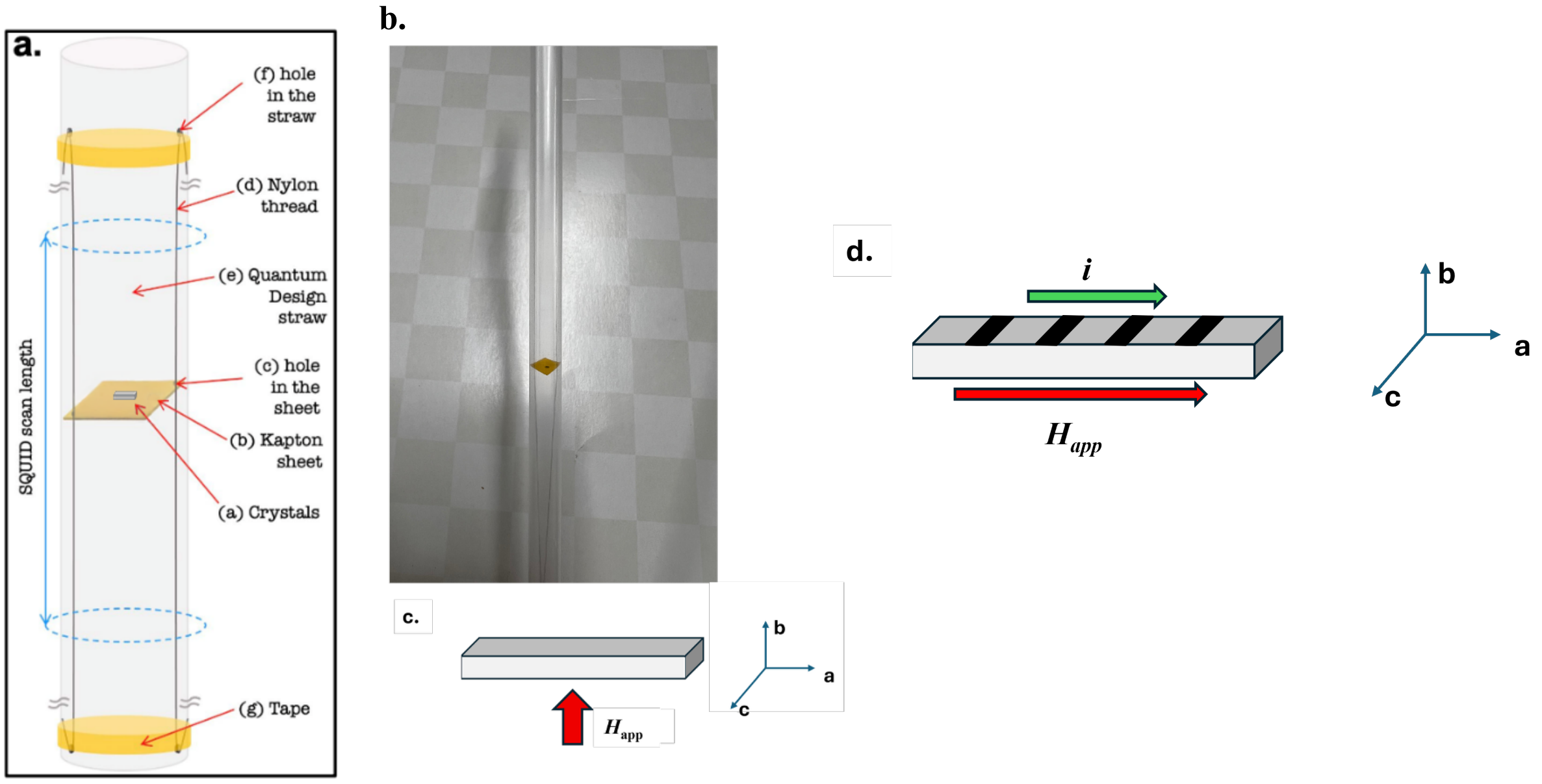

Figure. 1. (a, b) The sample holder illustration for SQUID measurement. The schematic of the needle-*like* crystal for (c) SQUID measurement with the applied magnetic field parallel to the b-axis, (d) magnetoresistance measurement with applied field parallel to the long a-axis. The lengths of the b- and c- are approximately equal while a-axis as the long axis. The ab-plane represent the conducting plane of the crystal.

thread, which was used to stabilize the sample position. Figure 1(c) shows the sample illustration that later was placed inside the straw, with the external applied magnetic field direction parallel to the b-axis, and hence parallel to the *ab* conducting plane. As for the crystal structure, we confirmed by using X-Ray diffractometer BRUKER D8-Quest at room temperature.

The magnetic measurements were performed to evaluate the superconducting transition temperature ($T_{c(diamagnetism)}$), superconducting volume fraction, and lower critical field ($H_{c1}$), as well as their evolution with increasing Br substitution. The superconducting transition temperature ($T_{c(diamagnetism)}$) was determined from the $\chi$-$T$ measurements. Since the lower critical field of organic superconductors is typically only several oersteds, residual magnetic fields trapped in the superconducting magnet can introduce a non-negligible systematic error in low-field magnetization measurements. Therefore, prior to each measurement, the superconducting magnet was reset (degaussed), and the magnetic field was carefully zeroed to minimize the remanent field. The actual magnetic field applied to the sample, $H_{act}$, was then used for the subsequent demagnetization correction. Moreover, by considering the demagnetization factor from the single crystal measurement, the effective field, $H_{eff}$, was obtained and represents the corrected field felt by the needle-*like* crystal. The demagnetization effect in CGS unit is denoted as in Eq. 3, with $N$ is a positive dimensionless demagnetizing factor [21]. The needle-*like* shape can be assumed has as a prolate spheroid shape with a-axis as the long axis and the b- and c-axes approximately equal. Since the applied magnetic field is oriented parallel to the b-axis, the diamagnetic response of the sample generates a magnetization opposite to the direction of $H_{app}$, and due to the sample's geometry, this magnetization produces a demagnetizing field that partially cancels the applied field, thereby reducing the effective internal field. Therefore, $H_{eff}$ is expected to be higher than $H_{act}$. By substituting the value of $N_b$ in Eq. 4 to the Eq.3 we obtained $H_{eff}$ for all the single crystal samples.

$$H_{eff} = H_{act} - 4\pi NM, \text{ with } H_{act} = H_{app} + H_{rem} \quad (3)$$

$$N_a + N_b + N_c = 1, \text{ where } N_b = N_c \quad (4)$$
$$N_a = \frac{1}{q^2}(\ln 2q - 1), \text{ with } q = {}^{a}/_{b}$$

where $N_a$, $N_b$, and $N_c$ denote the demagnetization factor in the field parallel to a-, b-, and c-axis. The $a/b$ ratio is calculated from the length ($a$- axis) and width ($b$-axis) of the crystal, as shown in Fig. 1(c). The $a/b$ ratio of $x_{c(rf)}$=0.22, 0.24; 0.42; and 0.63 are 1.44/0.18; 1.14/0.11; 0.54/0.16; and 1.10/0.14, respectively. Since the assumed sample geometry considers $b \approx c$, whereas the actual dimensions show a slight difference, the discrepancy between the b- and c- values was used to estimate the uncertainty in $N_b$. The resulting $N_b$ values are 0.486(12); 0.491(7); 0.48(2); and 0.486(11) for $x_{c(rf)}$=0.22, 0.24; 0.42; and 0.63, respectively.

Furthermore, the superconducting shielding fraction ($4\pi\chi_{int}$) from $\chi$-$T$ graph was calculated by using Eq. 5. Subsequently, the lower critical field ($H_{c1}$) was determined from $M$-$H$ measurements performed at $T < T_c$ with the external magnetic field swept from 0 to 200 Oe.

$$\chi_{int} = 4\pi \frac{\chi_{meas}}{1-4\pi N_b \chi_{meas}}, \text{ with } \chi_{meas} = \frac{M\ (emu/cm^3)}{H_{actual}} \quad (5)$$

Where $\chi_{int}$ represent demagnetization-corrected susceptibility or intrinsic magnetic response. Meanwhile $\chi_{meas}$ represent the susceptibility obtained directly from the measurement.

## 3. Results and Discussion

*3.1 Bromine concentration determination*

The accurate determination of Bromine content is essential to establish the chemical pressure-temperature phase diagram of the $\lambda$-Ga family. In previous study, the actual Br concentration which denotes as $x_c$ was estimated from the average expansion of the halogen bond lengths in the $GaBr_xCl_{4-x}$ tetrahedron obtained by single-crystal X-ray diffraction (SCXRD) [18]. Further refinement of the atomic position within the unit cell from the SCXRD has been reported in another study [20]. Although this approach successfully provided a quantitative estimation of the chemical composition ($x_c$), it was based on the averaged structural parameters rather than direct refinement of the halogen occupancies.

Recently, a full crystallographic refinement of the halogen occupancies was performed, providing a more reliable determination of the Br concentration. The refinement crystal structure demonstrates that the Br atoms statistically occupy similar site as Cl atoms within the $GaBr_xCl_{4-x}$ tetrahedron, with higher occupancies at the X2 and X4 sites [20]. These preferable sites are consistent with our previous bond-length expansion model to determine the Br concentration [18]. Throughout the present work, this refined Br composition is denoted as $x_{c(rf)}$ to distinguish it from the bond-length expansion method ($x_c$) reported previously.

As an independent compositional analysis, scanning electron microscope equipped with energy-dispersive X-ray spectroscopy (SEM–EDX) was also performed in our previous study, and the corresponding Br concentration was denoted as $x_{c(EDX)}$ [19]. Although the EDX analysis provided a semi-quantitative estimation of the Br concentration that was generally consistent with both the stoichiometry and $x_c$, its quantitative accuracy was limited as described in Section 2.1. Therefore, the EDX results were regarded as a qualitative confirmation of the compositional trend, whereas EPMA provided quantitative compositions after normalization to the corresponding standards. Table 1-5 summarize the quantitative EPMA results obtained after normalization to the standard samples (STD), as described in the Experimental section.

**Table 1.** Quantitative EPMA results for the $x$=0. Here, $\mu$ represents the average atomic percentage and $\mu_N$ is the composition normalized to Ga atom. The Cl and Br occupancies were calculated by using Eq. 1.

| No. of scan spots | Element | Line | $\mu$ | $\mu_N$ | Cl occupancy | Br occupancy | $\sum Cl, Br$ |
|---|---|---|---|---|---|---|---|
| 1 | C | K | 53.62 | 24.61 | 4 | 0 | 4 |
| | S | K | 17.19 | 7.89 | | | |
| | Se | L | 18.76 | 8.61 | | | |
| | Ga | K | 2.18 | 1.00 | | | |
| | Cl | K | 8.25 | 3.79 | | | |

**Table 2.** Quantitative EPMA results for the $x_{c(rf)}$=0.22. The quantities of $\mu$, $\mu_N$ and the Cl/Br occupancies are defined as in Table 1.

| No. of scan spots | Element | Line | $\mu$ | $\mu_N$ | Cl occupancy | Br occupancy | $\sum Cl, Br$ |
|---|---|---|---|---|---|---|---|
| 11 | C | K | 73.34 | 58.81 | 3.802 | 0.198 | 4 |
| | S | K | 12.53 | 10.05 | | | |
| | Se | L | 12.44 | 9.98 | | | |
| | Ga | K | 1.25 | 1.00 | | | |
| | Cl | K | 0.42 | 0.34 | | | |
| | Br | L | 0.02 | 0.02 | | | |

**Table 3.** Quantitative EPMA results for the $x_{c(rf)}$=0.24. The quantities of $\mu$, $\mu_N$ and the Cl/Br occupancies are defined as in Table 1.

| No. of scan spots | Element | Line | $\mu$ | $\mu_N$ | Cl occupancy | Br occupancy | $\sum Cl, Br$ |
|---|---|---|---|---|---|---|---|
| 4 | C | K | 71.59 | 39.13 | 3.76 | 0.24 | 4 |
| | S | K | 12.31 | 6.73 | | | |
| | Se | L | 11.64 | 6.36 | | | |
| | Ga | K | 1.83 | 1.00 | | | |
| | Cl | K | 2.48 | 1.35 | | | |
| | Br | L | 0.16 | 0.09 | | | |

**Table 4.** Quantitative EPMA results for the $x_{c(rf)}$=0.42. The quantities of $\mu$, $\mu_N$ and the Cl/Br occupancies are defined as in Table 1.

| No. of scan spots | Element | Line | $\mu$ | $\mu_N$ | Cl occupancy | Br occupancy | $\sum Cl, Br$ |
|---|---|---|---|---|---|---|---|
| 4 | C | K | 86.41 | 135.5 | 3.56 | 0.44 | 4 |
| | S | K | 5.55 | 8.70 | | | |
| | Se | L | 6.17 | 9.67 | | | |
| | Ga | K | 0.64 | 1.00 | | | |
| | Cl | K | 1.11 | 1.73 | | | |
| | Br | L | 0.14 | 0.21 | | | |

**Table 5.** Quantitative EPMA results for the $x_{c(rf)}$=0.63. The quantities of $\mu$, $\mu_N$ and the Cl/Br occupancies are defined as in Table 1.

| No. of scan spots | Element | Line | $\mu$ | $\mu_N$ | Cl occupancy | Br occupancy | $\sum Cl, Br$ |
|---|---|---|---|---|---|---|---|
| 2 | C | K | 66.92 | 42.12 | 3.38 | 0.62 | 4 |
| | S | K | 13.97 | 8.79 | | | |
| | Se | L | 16.78 | 10.56 | | | |
| | Ga | K | 1.59 | 1.00 | | | |
| | Cl | K | 0.62 | 0.39 | | | |
| | Br | L | 0.11 | 0.07 | | | |

From the Table 1, the measured atomic percentages of S, Se, and Cl are approximately twice that of Ga, consistent with the expected stoichiometry of $\lambda$-$(BETS)_2GaCl_4$. Therefore, the elemental composition was normalized to the Ga atom to obtain the chemical formula. However, the atomic percentage of Carbon is less reliable because of the limited analytical accuracy for light elements arising from their low-energy characteristic x-rays. The Br and Cl occupancies were subsequently calculated using Eq. (2), and the resulting Br concentrations agree well with those obtained from the SCXRD refinement, $x_{c(rf)}$. The Br concentrations determined by the different characterization techniques are summarized in Table 6. The good agreement between the EPMA and SCXRD refinement results confirms the reliability of $x_{c(rf)}$. Therefore, the refined Br compositions, $x_{c(rf)}$, are hereafter adopted as the representative Br concentrations for the samples $x$=0.2 S1, 0.2 S2, 0.4S1, and 0.65. These values are used throughout the following discussion of the magnetic properties, as well as in the construction of the pressure–temperature ($P$–$T$) and field–pressure ($H$-$P$) phase diagrams. The mapping results of the single crystals samples for $x_{c(rf)}$=0.24 and 0.63 are presented in Figure 2.

**Table 6.** Summary of the determination of Br concentration from several techniques.

| Stoichiometry | XRD | | SEM-EDX | EPMA |
|---|---|---|---|---|
| $x$ | $x_c$ | $x_{c(rf)}$ | $x_{c(EDX)}$ | $x_{c(EPMA)}$ |
| 0.2S1 | 0.197(37) | 0.22 | 0.21(2) | 0.198 |
| 0.2S2 | 0.193(56) | 0.24 | 0.25(2) | 0.24 |
| 0.4S1 | 0.349(50) | 0.42 | 0.46(2) | 0.44 |
| 0.65 | 0.531(50) | 0.63 | 0.56(3) | 0.62 |

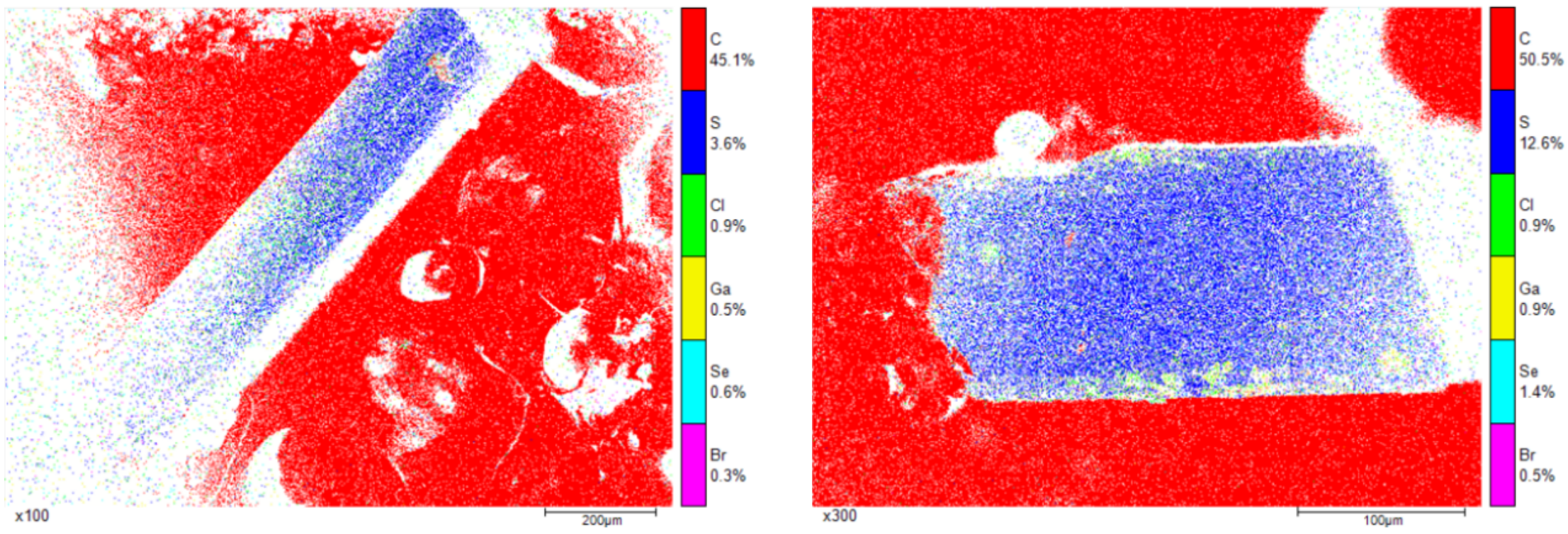

Figure. 2. The mapping analysis of the (a) $x_{c(rf)}$=0.24 and (b) $x_{c(rf)}$=0.63.

### *3.2 Superconducting magnetic properties of $\lambda$-$(BETS)_2GaBr_xCl_{4-x}$*

Figure 3 shows the temperature dependence of the demagnetization-corrected magnetic susceptibility measured with the magnetic field applied parallel to the b-axis as shown in the schematic Fig. 1(c). All magnetic data presented in this section have been corrected for the demagnetization effect using the procedure described in the Experimental section. The field felt by the single crystal sample represent by the $H_{eff}$ and was calculated by using Eq. 3. The right axis represents the corresponding superconducting shielding fraction calculated from the internal susceptibility using Eq. 5. The superconducting transition temperature, $T_{c(diamagnetism)}$, was determined from the onset of the diamagnetic response, as indicated by the arrows in Fig. 3. A large diamagnetic response owing to the superconductivity was observed below $T_{c(diamagnetism)}$. The onset temperature of this response agrees well with the zero-resistance temperature reported previously [18]. As the Br concentration increases, $T_{c(diamagnetism)}$ also increasing consistent with the resistivity measurements of the mixed crystal reported by Tanaka *et al.* [15].

For all samples, the field-cooled (FC) susceptibility is smaller than the zero-field-cooled (ZFC) susceptibility, reflecting irreversible magnetic behavior associated with magnetic flux trapping and vortex pinning after flux penetration into the superconducting state. Such behaviour is commonly observed in type-II superconductors and is consistent with the *M*-*H* results as discussed below. By comparing the $x_{c(rf)}$=0.22 and 0.24, the gradient of the $x_{c(rf)}$=0.22 appear to be steeper, indicating that the superconducting state is less robust compared to $x_{c(rf)}$=0.24. This behavior was consistent with the resistivity result, in which for $x_{c(rf)}$=0.22 exhibits the broad upturn anomaly below 10K which may indicates the enhanced of the localized electron at low Br concentration [18]. The robustness of the superconducting state for $x_{c(rf)}$=0.24 was further supported by the high value of the $H_{c1}$ as shown in Fig. 6 as will be discussed later. Nevertheless, both samples exhibit nearly identical superconducting transition temperatures, $T_{c(diamagnetism)}$, of 5.6(2) K and 5.6(4) K for $x_{c(rf)}$=0.22 and 0.24, respectively. Furthermore, for $x_{c(rf)}$=0.42, the superconducting shielding fraction exceeds the full shielding diamagnetism ($4\pi\chi_{int} = -1$ ). Meanwhile the obtained $H_{c1}$ from the Fig. 6(c) shows about 4.5(4) Oe in which slightly below the $H_{eff}$ of $x_{c(rf)}$=0.42, therefore the exceedingly more than 100% of superconducting shield fraction likely not simply due to the error in the applied magnetic field caused by the residual magnetic field during the measurement but also attributed to uncertainty in the demagnetization correction because its sensitivity to the actual crystal geometry. Although the exact origin cannot be identified from the present measurement, the observation of superconducting volume fraction for the following sections is evaluated from the low-field *M*-*H* measurement at 2K as plotted by the blue triangular symbols in Fig. 3, where the Meissner response can be evaluated more accurately. The diamagnetic volume fraction obtained from the $M$-$H$ measurements are approximately 73.32(6)%, 74.3(1)%, 96.1(9)%, and 30.5(1)% for $x_{c(rf)}$=0.22, 0.24, 0.42, and 0.63, respectively. The relatively large shielding fractions observed for $x_{c(rf)}$=0.22, 0.24, and 0.42 indicate that these samples exhibit bulk superconductivity. However, the suppression of diamagnetic volume fraction for $x_{c(rf)}$=0.63 is consistent with the pronounced resistivity upturn anomaly below 10 K as the system approaches the Mott transition [18], suggesting an increase in the non-superconducting volume fraction. Such behavior is also consistent with the proximity of the spin-density-wave (SDW) phase reported for $x$=0.75 [22].

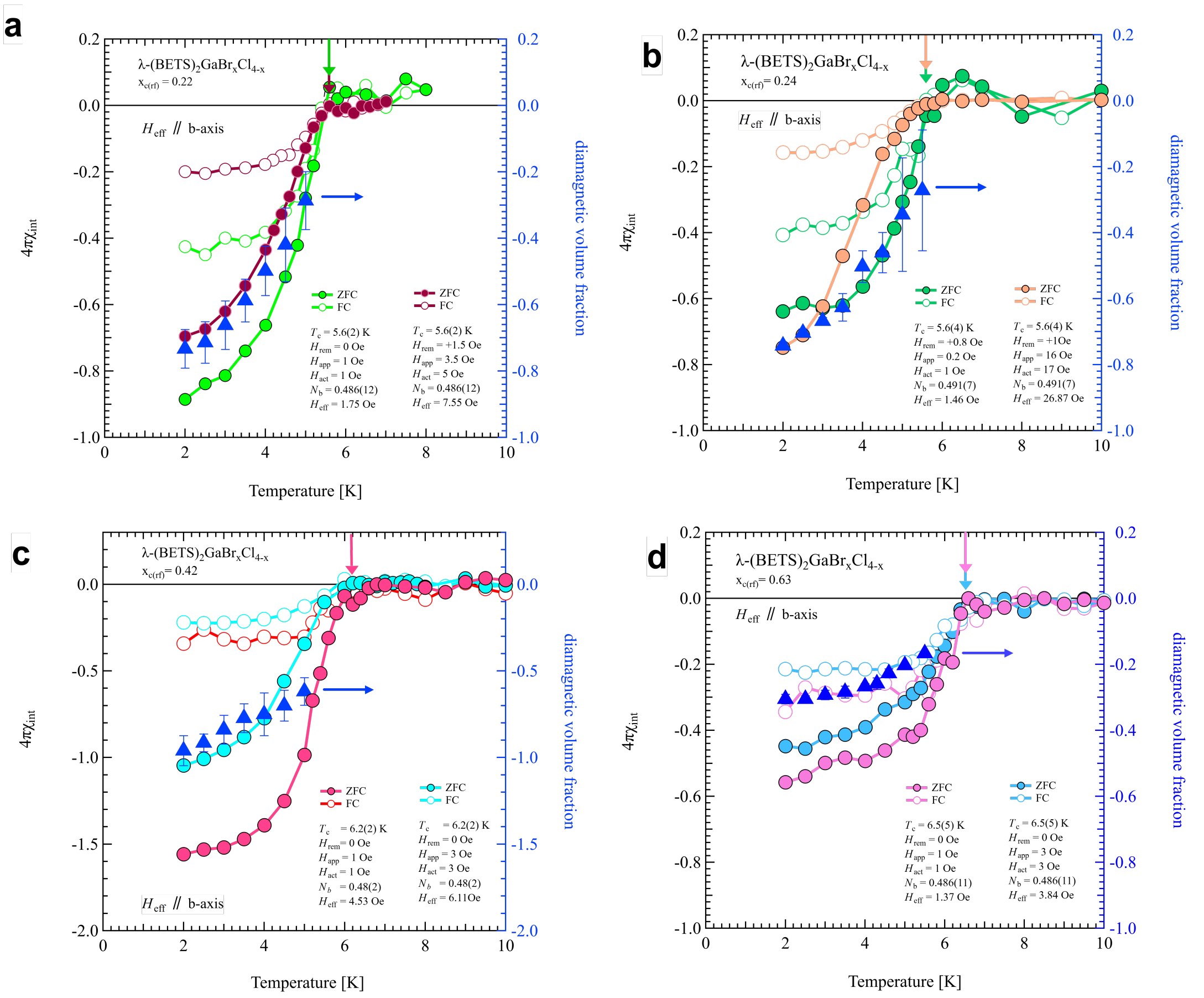

Figure. 3. Temperature dependence of demagnetization-corrected magnetic susceptibility, $4\pi\chi_{int}$, measured with the magnetic field applied parallel to the b-axis for (a) $x_{c(rf)}$=0.22, (b) $x_{c(rf)}$=0.24, (c) $x_{c(rf)}$=0.42, (d) $x_{c(rf)}$=0.63. The uncertainty of the $N_b$ was calculated by the deviation of the actual crystal dimensions from the assumed geometry. The triangular symbols denote the normalized diamagnetic volume fraction derived from the slope of the corresponding $M$–$H$ curves at several temperature, $T<T_c$. The error bars represent the uncertainty obtained from the linear fitting of the $M$–$H$ data.

To further investigate the superconducting state, the $M$-$H$ measurements were performed below $T_c$. Besides providing a more reliable evaluation of the superconducting volume fraction, the $M$-$H$ curves also enable determination of the lower critical field, $H_{c1}$, through analysis of the deviation from the linear Meissner response. Figure 4 presents the magnetic field dependence of the magnetization measured at several temperatures below $T_c$ for all investigated Br concentrations. At low magnetic fields, all samples exhibit a linear magnetization characteristic of the Meissner state. As the applied field increases, the magnetization gradually deviates from linearity, indicating the onset of magnetic flux penetration. The deviation from liner fitting at low field was defined as lower critical field, $H_{c1}$. To determine $H_{c1}$ more precisely, the square root of the magnetization deviation from the Meissner line, $\sqrt{\Delta M}$, was plotted as a function of the effective magnetic field, as shown in Fig. 5. Here, $\Delta M = M_{\mathrm{meas}} - M_{\mathrm{fit}}$, where $M_{\mathrm{fit}}$ is the linear fit to the low-field Meissner region. According to the Bean critical-state model, the initial flux penetration into a type-II superconductor follows $\Delta M^{\frac{1}{2}} \propto (M_{meas} - M_{fit})^{\frac{1}{2}}$ allowing $H_{c1}$ at each temperature to be obtained by extrapolating the linear line of the $\sqrt{\Delta M}$ plot to zero. The temperature dependence of the lower critical field obtained from this procedure is presented in Fig. 6. The experimental data were then fitted using the two-fluid model, $H_{c1}(T)=H_{c1}(0)\,[1-(T/T_c)^2]$ from which the zero-temperature lower critical field, $H_{c1}(0)$, was determined [23].

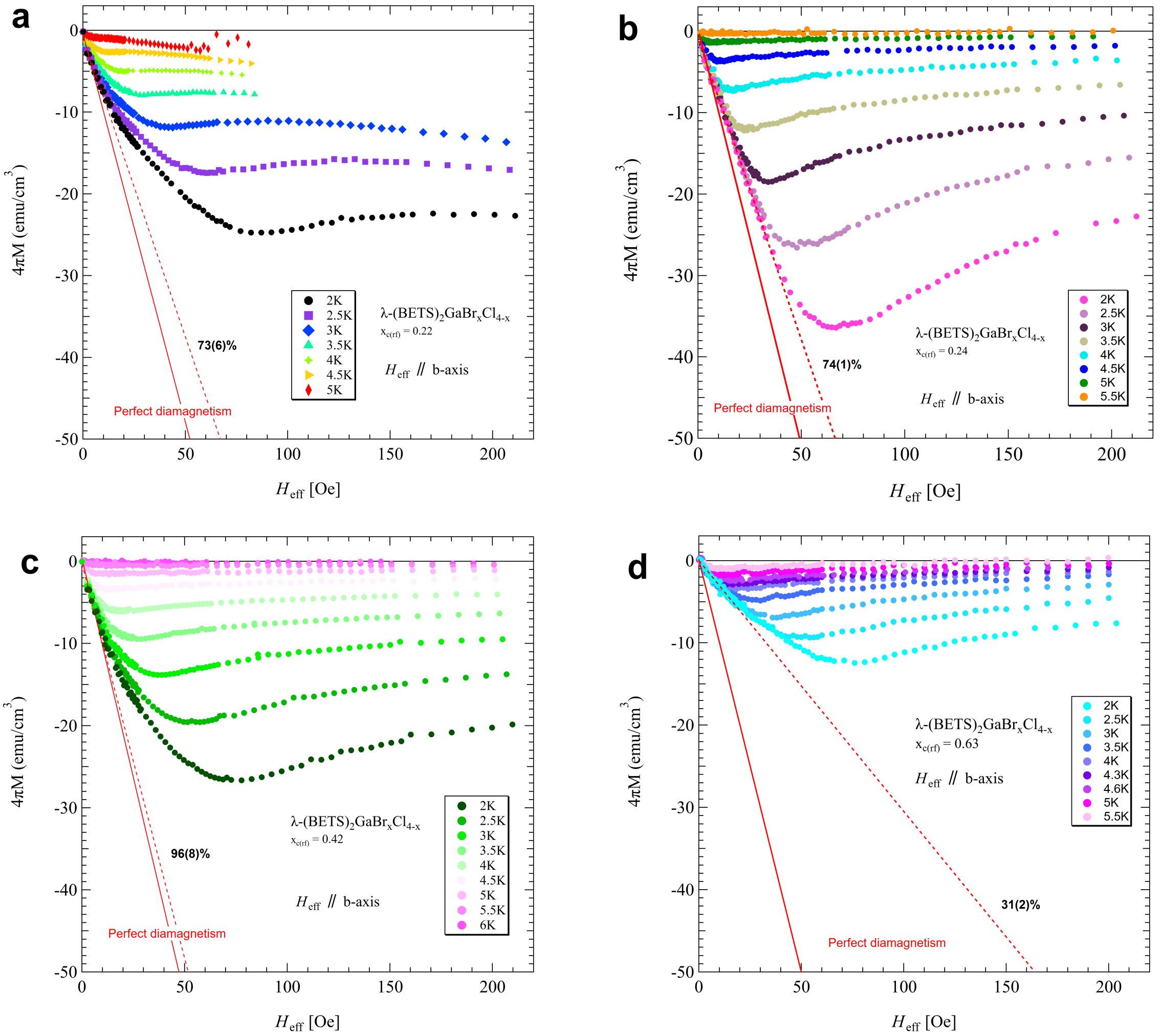

Figure. 4. The field dependence of the magnetization, $4\pi M(emu/cm^3)$, measured with the magnetic field applied parallel to the b-axis at several temperatures below $T_c$ , for (a) $x_{c(rf)}$=0.22, (b) $x_{c(rf)}$=0.24, (c) $x_{c(rf)}$=0.42, (d) $x_{c(rf)}$=0.63. The solid red line indicates the ideal diamagnetic response corresponding to a superconducting volume fraction of 100%, whereas the red dashed line is a linear fit to the low-field *M*-*H* data at 2 K. The lower critical field, $H_{c1}$, was determined from the deviation of the *M*-*H* curve from this linear behavior.

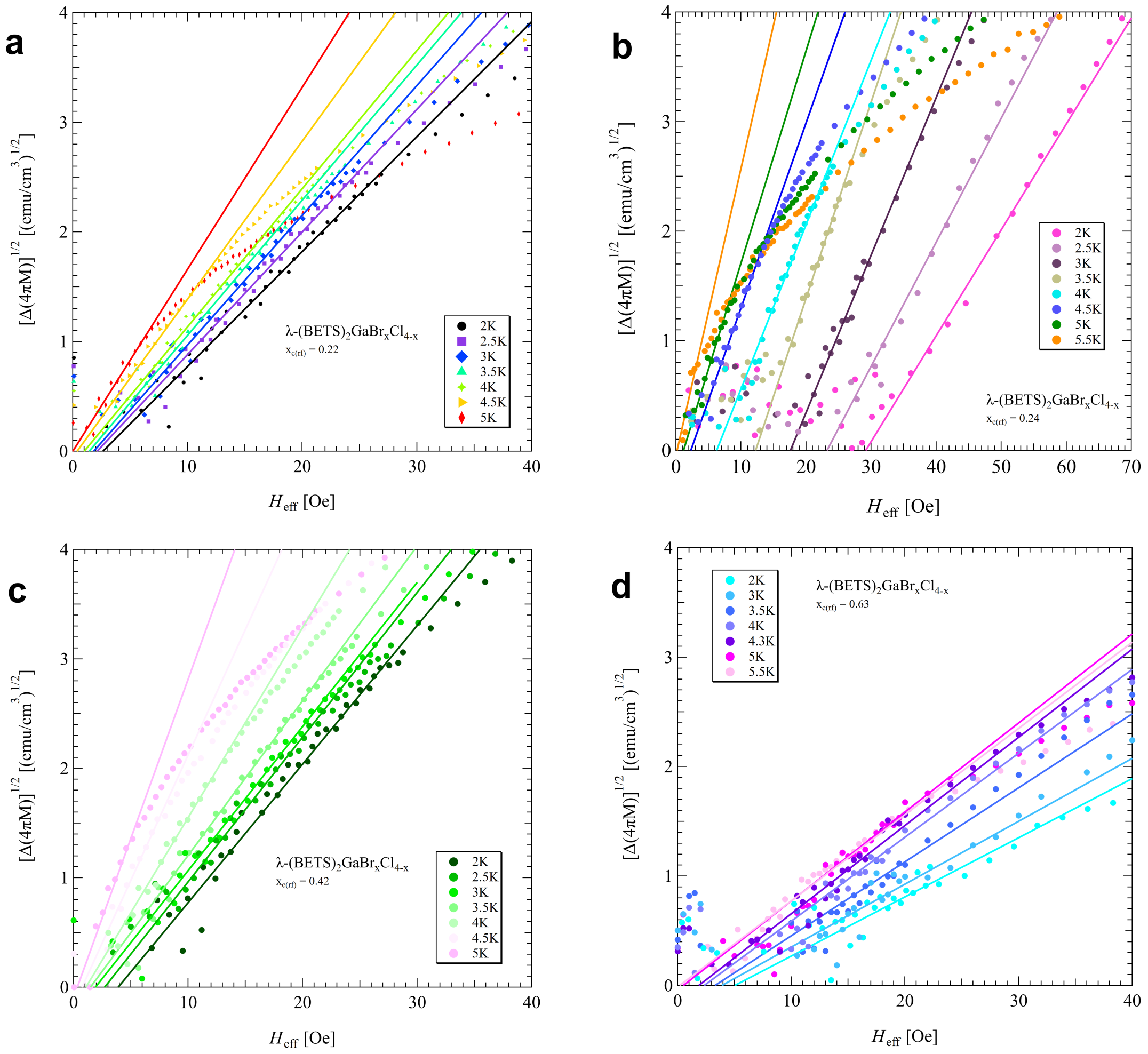

Figure. 5. Square root of the deviation from the linear Meissner response, $[\Delta(4\pi M)]^{1/2}$, as a function of the demagnetization-corrected effective magnetic field, $H_{\mathrm{eff}}$, at several temperatures below $T_c$ for (a) $x_{c(rf)}$=0.22, (b) $x_{c(rf)}$=0.24, (c) $x_{c(rf)}$=0.42, (d) $x_{c(rf)}$=0.63. The deviation was calculated as $\Delta(4\pi M) = 4\pi M_{\mathrm{meas}} - 4\pi M_{\mathrm{fit}}$, where $M_{\mathrm{fit}} = aH_{\mathrm{eff}} + b$ represents a linear fit to the low-field Meissner response. The lower critical field, $H_{c1}$, was determined by extrapolating the linear fits to $[\Delta(4\pi M)]^{1/2} = 0$.

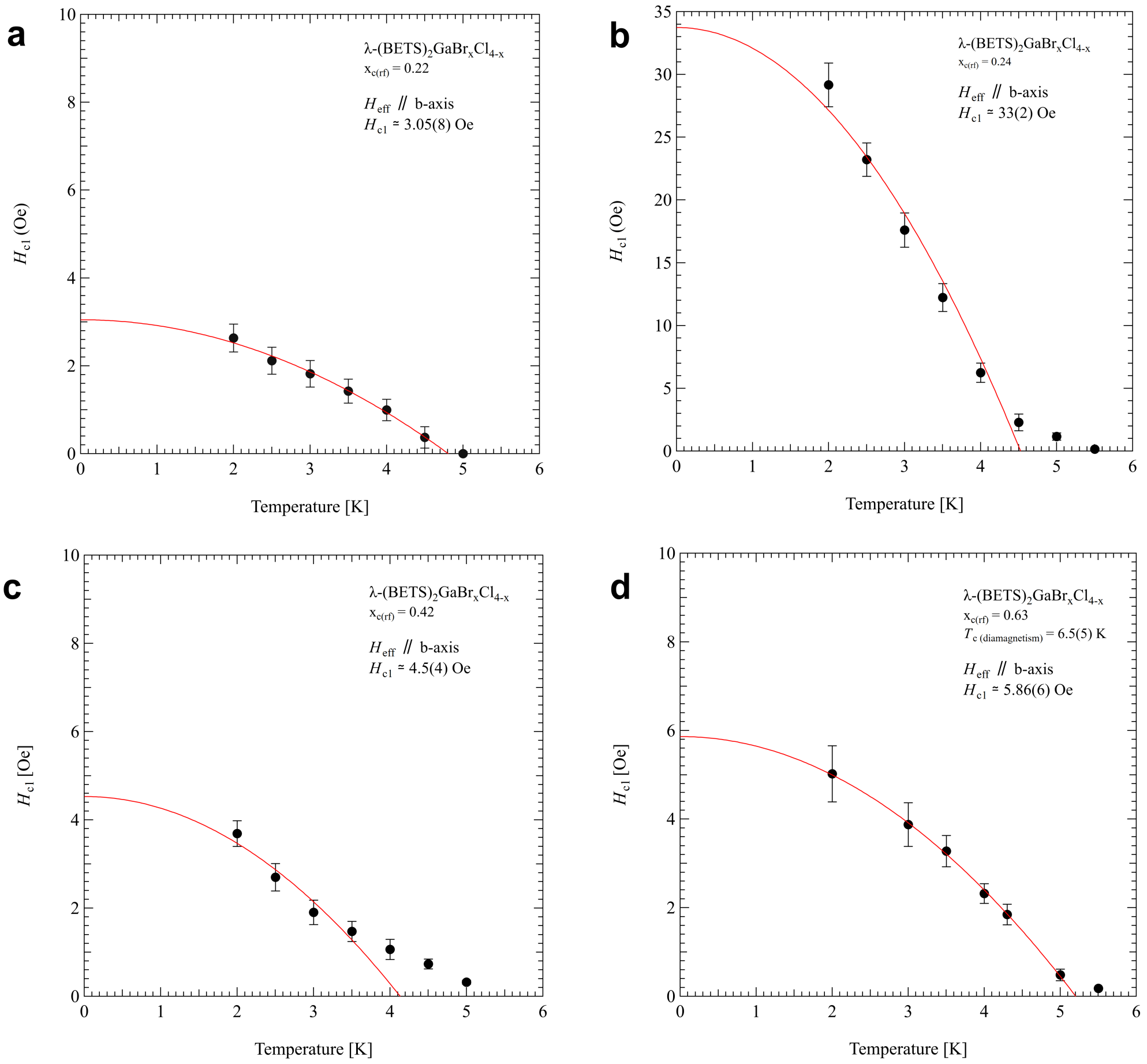

Figure. 6. Temperature dependence of the lower critical field, $H_{c1}$ , including the error bars and after the correction for the demagnetization factor $N_c$, for (a) $x_{c(rf)}$=0.22, (b) $x_{c(rf)}$=0.24, (c) $x_{c(rf)}$=0.42, (d) $x_{c(rf)}$=0.63. The red solid line represents the fitting result of the data using two-fluid model equation, $H_{c1}$(T)=$H_{c1}$(0) [1-($T$/$T_c$)2].

Figure 6 represent the temperature dependence of the lower critical field, $H_{c1}$, for all investigated Br concentrations. The $H_{c1}$ decreases monotonically with increasing temperature and approaches zero near $T_c$ for all the mixed crystal samples, as expected for a type-II superconductor. The solid curve in which represent the fitting using two-fluid model reproduce the experimental data well over the measured temperature range. The obtained $H_{c1}$ values exhibit a nonmonotonic dependence on Br substitution, with a pronounced maximum at $x_{c(rf)}$=0.24 compared with the other $\lambda$-Ga compounds. Although $T_{c(diamagnetism)}$ and the superconducting volume fraction are similar to $x_{c(rf)}$=0.22, the superconducting state itself responds differently to magnetic field. The difference in $H_{c1}$ therefore suggests that the superconducting state is intrinsically more robust against magnetic flux penetration for $x_{c(rf)}$=0.24. Such behavior is consistent with the anomalous transport properties previously reported in the low-Br region at $x_{c(rf)}$=0.22 as discussed above [18], although the microscopic origin of this $H_{c1}$ enhancement remains unclear. Further local magnetic probes, such as μSR, would be valuable for clarifying the origin of this anomalous magnetic response. However, as shown by the EPMA and SCXRD analyses, crystals grown from the same stoichiometry composition ($x$=0.2) exhibit slight variations in the refined Br concentration, resulting in the differences of the transport and magnetic properties. Such sensitivity to the actual Br content makes microscopic investigations using local magnetic probes become experimentally challenging because multiple crystals with nearly identical Br distributions and reproducible superconducting properties would be required for meaningful microscopic comparisons. Nonetheless, the difference of electron transport and Meissner state properties between $x_{c(rf)}$=0.22 and 0.24 indicate a characteristic of superconducting phase boundary within low Br concentration region. Notably, the $H_{c1}$ evolution exhibits a dome-*like* structure with maximum at $x_{c(rf)}$=0.24.

Furthermore, the upper critical field, $H_{c2}$, was plotted together with $H_{c1}$ in the $H$–$P$ phase diagram shown in Fig. 9. Both quantities were measured with the magnetic field applied within the ab conducting plane, with $H \parallel b$ for the $H_{c1}$ measurements and $H \parallel a$ for the $H_{c2}$ measurements. Accordingly, the $H$–$P$ diagram is intended to represent the evolution of the characteristic superconducting field scales under chemical pressure, rather than a direct comparison of $H_{c1}$ and $H_{c2}$ for an identical crystallographic field orientation. The $H_{c2}$ values were determined using several resistive criteria: 0.1%, 50%, and 90%, corresponding to the nearly zero-resistance, midpoint, and onset regions of the resistive transition, respectively, at each applied magnetic field. However, since the $H_{c2}$ value for $x$=0, taken from Ref. [28], was determined using the midpoint criterion with the magnetic field applied parallel to the conducting plane, the midpoint values for $x_{c(rf)}$=0.22 and 0.42 were adopted in Fig. 9 for consistent comparison. The midpoint criterion was defined as the temperature at which the resistivity reached 50% of the normal-state resistivity just above the superconducting transition. Accordingly, for samples $x_{c(rf)}$=0.22 and 0.42, the initial slope of $H_{c2}$(T) near $T_c$ was then used to estimate the zero-temperature upper critical field using the phenomenological Ginzburg–Landau expression, $H_{c2}(0) = -0.5T_c \left( {dH_{c2}}/{dT} \right)_{T_c}$ [29].

Meanwhile, for $x_{c(rf)}$=0.63, the obtained $H_{c2}$ was taken from Ref [30] using the same single crystal of the sample stoichiometry, $x$=0.65. The value of $H_{c2}$ was estimated to be approximately 15 T from the disappearance of the superconducting transition under applied magnetic field parallel to the conducting plane. The magnetoresistance result including the $H_{c2}$-$T$ plot of $x_{c(rf)}$=0.22 and 0.42 represented in Fig. 7 and 8, respectively. The resulting $H_{c2}$ values for $x_{c(rf)}$=0.22 and 0.42 systematically increase as the resistive criterion is raised from 0.1% to 50% and 90%, consistent with the expected evolution from the nearly zero-resistance to the onset region of the superconducting transition.

Within the anisotropic Ginzburg–Landau (GL) framework, the effective coherence length for a magnetic field applied parallel to the conducting plane is defined as the geometric mean of the coherence lengths perpendicular and parallel to the plane $\xi_{eff} = \sqrt{\xi_{\perp}\xi_{\parallel}} = \sqrt{\frac{\Phi_0}{2\pi H_{c2}^{\parallel}}}$. Using the midpoint criterion for $H_{c2}$, the estimated $\xi_{eff}$ values for $x_{c(rf)}$=0.22, 0.42 and 0.63 are approximately 5.37, 7.67, 4.68 nm, respectively. A larger $\xi_{eff}$ represents a longer characteristic length scale over which the superconducting order parameter varies spatially or recovers from a local perturbation. The resulting nonmonotonic variation of $\xi_{eff}$ provides an initial indication that the characteristic superconducting length scale evolves across the chemical-pressure region. However, because the available $H_{c2}$ values are obtained from different measurements and compositions, this trend should presently be regarded as phenomenological.

For a quasi-two-dimensional superconductor, application of the magnetic field parallel to the conducting plane is generally expected to reduce the orbital pair-breaking contribution compared with the perpendicular-field configuration. To assess the possible relevance of paramagnetic pair breaking, the weak-coupling Pauli limiting field was estimated using $\mu_0 H_P = 1.86T_c$ [29]. Based on the midpoint transition temperatures, the resulting $H_P$ values are approximately 11.09 and 13.39 T for $x_{c(rf)}$=0.22 and 0.42, respectively. For $x_{c(rf)}$=0.63, $H_P$ was estimated to be approximately 12.09 T using a superconducting transition temperature of approximately 6.5 K [30]. The $H_{c2}$ values for $x_{c(rf)}$=0.22 and 0.63 are comparable to or slightly exceed the corresponding weak-coupling Pauli-limit estimates. However, this comparison alone is insufficient to determine the relative contributions of orbital and paramagnetic pair breaking. Measurements for both field orientations, together with an evaluation of the orbital limiting field and Maki parameter, are required to clarify the relative contributions of orbital and paramagnetic pair breaking. Such investigations are particularly relevant in view of the unconventional high-field superconducting behavior previously reported for pristine λ-$(BETS)_2GaCl_4$ [31]. The present $H_{c1}^{H\parallel b}$ and $H_{c2}^{H\parallel a}$ measurements therefore provide an initial $H$–$P$ description of the superconducting state on the negative chemical-pressure side of the λ-Ga phase diagram, complementing the broader $P$–$T$ phase diagram discussed in the following section. Extension of $H_{c2}$ measurements toward the physical-pressure side will be important for determining how the field robustness of superconductivity evolves across the broader pressure-tuned phase diagram. The refined Br concentrations obtained from SCXRD ($x_{c(rf)}$) were converted into chemical pressure using the experimentally determined bulk modulus relation, allowing the evolution of the superconducting magnetic properties to be discussed on the same pressure scale as the pressure-temperature ($P$-$T$) phase diagram presented in the following Section 3.3.

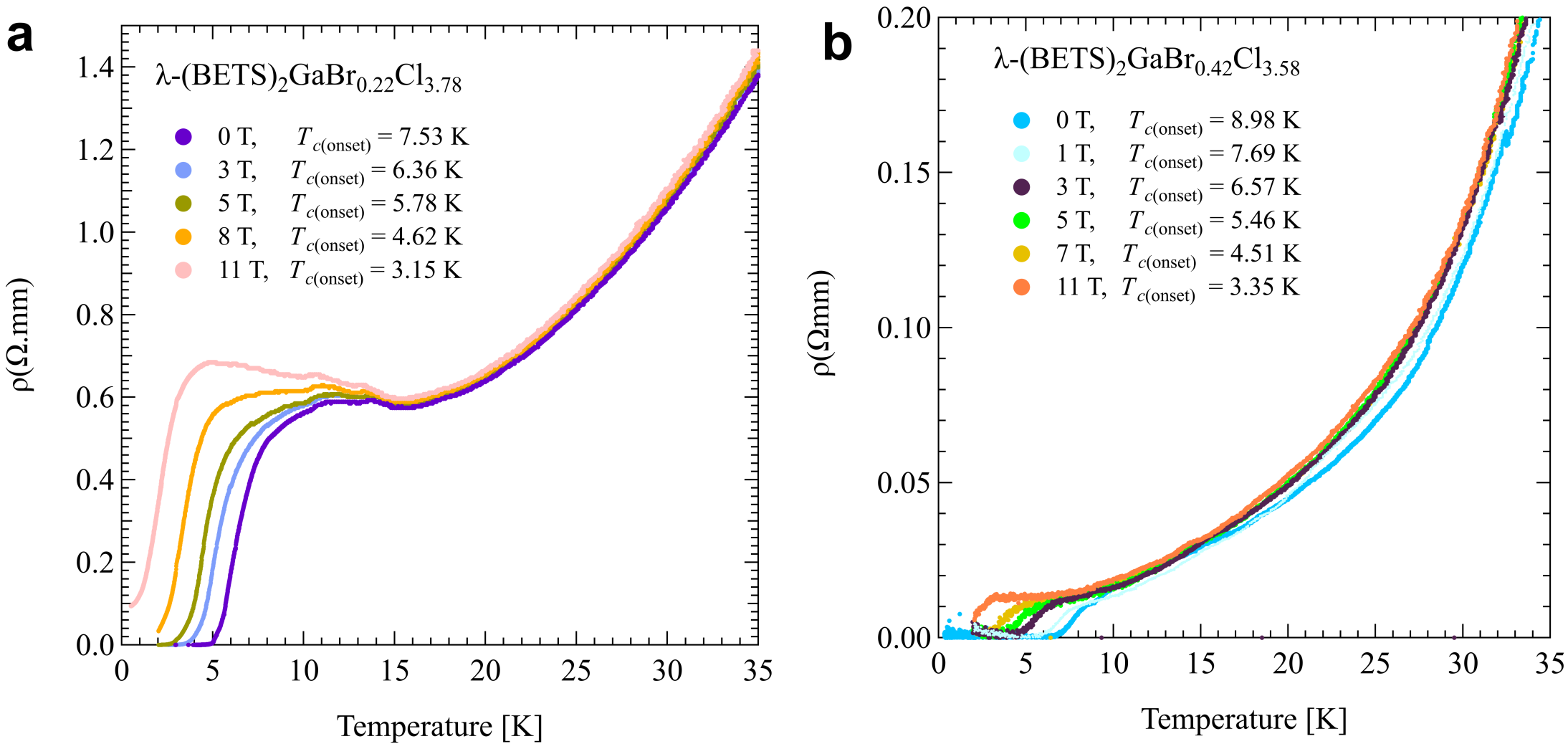

Figure. 7. Magnetoresistance measured under magnetic fields from 0 to 11 T applied parallel to the long a-axis (within the ab conducting plane). The data were used to construct the $H$–$T$ phase diagrams shown in Fig. 8, following the procedure of Ref. [28].

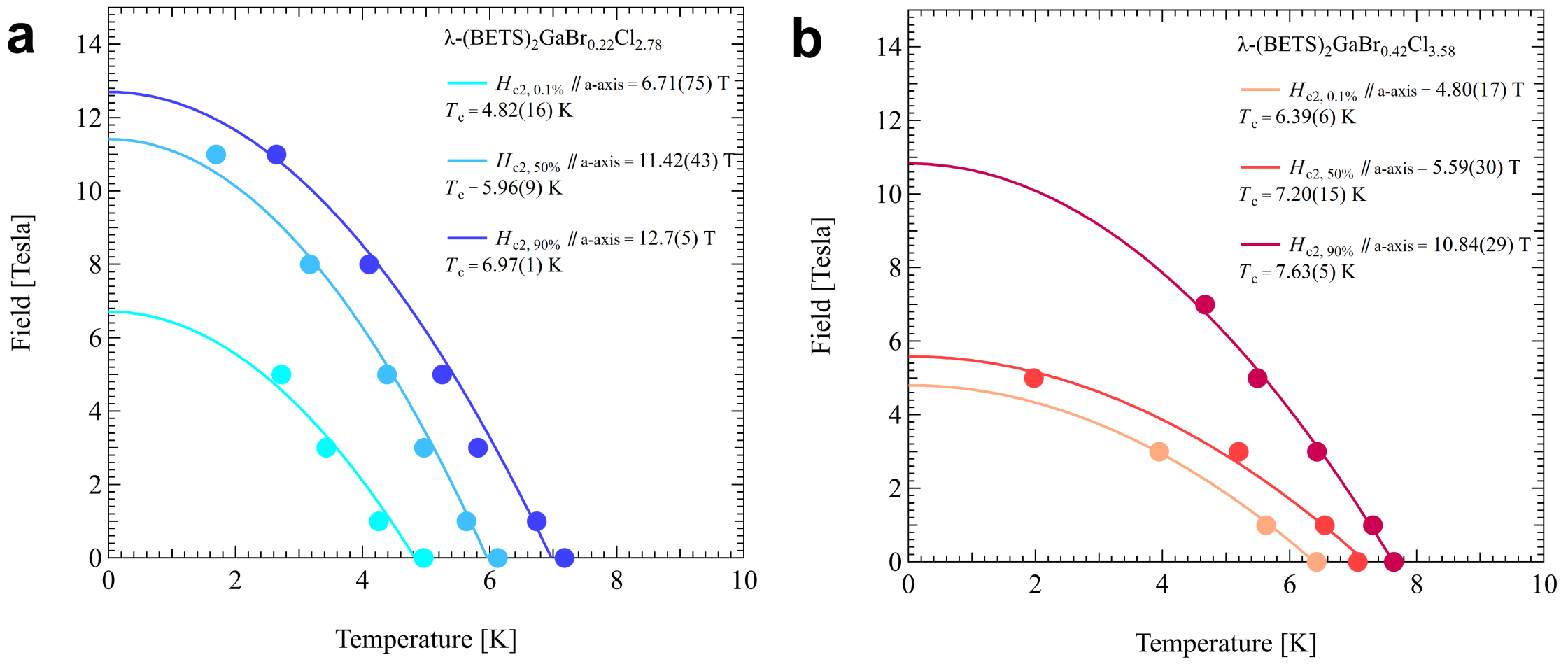

Figure. 8. Temperature dependence of the upper critical field, $H_{c2}$, for (a) $x_{c(rf)}$=0.22 and (b) $x_{c(rf)}$=0.42, determined using the nearly zero-resistance, midpoint, and onset criteria of the resistive transition shown in Fig. 7.

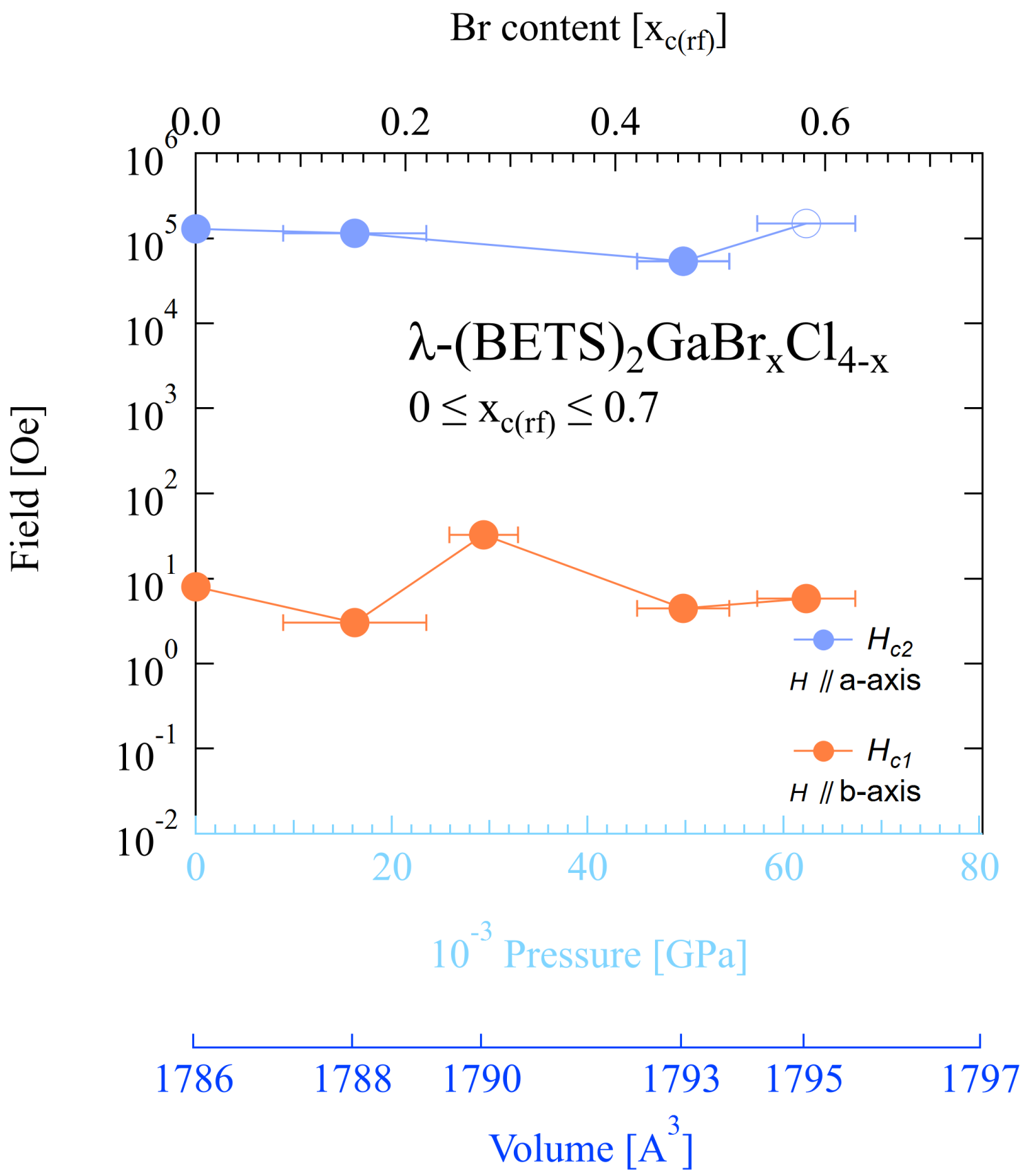

Figure. 9. Pressure dependence of the characteristic superconducting fields, $H_{c1}^{H\parallel b}$ and $H_{c2}^{H\parallel a}$ of λ-(BETS)$_2$GaBr$_x$Cl$_{4-x}$ (0≤x<0.75). The $H_{c2}$ value was taken from the midpoint resistive criterion in Figure. 8. The upper, middle, and lower horizontal axes represent the Br concentration, equivalent chemical pressure, and unit-cell volume, respectively. The open circle of $H_{c2}$ at $x_{c(rf)}$=0.63 was obtained from another magnetoresistance study reported in Ref. [30].

*3.3 Superconducting phase diagram of λ-Ga under chemical and physical pressure*

Combining the present magnetic data with previously reported transport data [18], we constructed an electronic *P–T* phase diagram for the λ-Ga system, shown in Fig. 10. Because Br substitution expands the unit-cell volume and drives the system toward the Mott-insulating state, the mixed-crystal series is placed on the negative-pressure side of λ-$(BETS)_2GaCl_4$. This behavior contrasts with κ-$(ET)_2Cu[N(CN)_2]Br_xCl_{1-x}$, where Br substitution instead increases the intermolecular transfer integrals and reduces $U/W$ [9] — i.e., acts as positive chemical pressure because the same effect achieved in κ-$(ET)_2Cu[N(CN)_2]Cl$ by applying ~0.3 kbar of physical pressure [9]. Br substitution therefore has opposite signs of chemical-pressure effect in the κ-$(ET)_2X$ and λ-Ga families.

A central feature of the proposed *P–T* phase diagram is a double-dome superconducting region with a local minimum in $T_c$ near $x$=0. We attribute this suppression to competing charge and antiferromagnetic (AF) fluctuations that develop near $x$=0, as evidenced by optical-conductivity [24] and $^{13}$C NMR [25] measurements, and that are further enhanced by hydrostatic pressure up to ~0.3 GPa [26]; these AF fluctuations are discussed further below in connection with the ~40 K transport crossover. This local suppression occurs in the pressure region where quantum-critical point (QCP) has previously been proposed at $P$≈0.13 GPa [14]— separating a low-pressure regime in which superconductivity coexists with magnetism and unconventional metallic state from a higher-pressure regime, where superconductivity is followed by conventional metallic behavior. This picture is consistent with the distinct superconducting behavior observed at low Br concentration ($x_{c(rf)}$=0.22 and 0.24; Section 3.2), where competing fluctuations may weaken the superconducting state and locally suppress $T_c$.

At higher Br concentration, $^{13}$C NMR measurements at $x$=0.75 identify a commensurate spin-density-wave (SDW) state, based on the similarity of $1/T_1T$ below 13 K to that of $(TMTTF)_2Br$ [22]. Our resistivity and magnetic data at $x_{c(rf)}$ = 0.63 — an enhanced non-superconducting volume fraction together with a pronounced low-temperature resistivity upturn [18] — are consistent with this composition lying close to the same SDW instability. The associated SDW fluctuations are gradually suppressed as the Br concentration decreases, and the SDW state itself is suppressed by physical pressure above ~0.3 GPa in λ-Ga [14,26] (green region, Fig. 10). Notably, the same broad low-temperature resistivity upturn anomaly is not confined to the chemical-pressure (Br-substitution) axis: it also appears in λ-Ga under physical pressure [14] and, via the pressure correspondence established below, in λ-STF under physical pressure [27]. This recurrence across three independent tuning routes — Br substitution, physical pressure in λ-Ga, and physical pressure in λ-STF — suggests that the SDW fluctuations may extend continuously from the chemical-pressure axis into the physical-pressure axis, rather than being confined to low Br concentration.

Below a crossover temperature of ~40 K, $x_{c(rf)}$=0.24 and 0.42 show comparable resistivities, roughly one order of magnitude higher than $x$=0, whereas $x_{c(rf)}$=0.22 is roughly two orders of magnitude higher. This nonmonotonic trend points to a crossover between two distinct transport regimes, represented in Fig. 10 by the yellow (lower-resistivity) and light blue (higher-resistivity) regions, with $x_{c(rf)}$=0.22 sitting at a local resistivity maximum near their boundary — consistent with a superconducting phase boundary near $x$≈0.2. Interestingly, previous $^{13}$C NMR measurements revealed antiferromagnetic (AF)-fluctuation in both pristine $\lambda$-Ga and $\lambda$-$(BETS)_2GaBr_{0.75}Cl_{3.25}$, which show kinks in $1/T_1T$ at ~55 K [25] and ~30 K, respectively [22]. This interpretation is further supported by theoretical calculations for $\lambda$-Ga which reproduce a comparable characteristic temperature, $T_{\mathrm{kink}} \approx 5$ meV (~ 58 K) [32]. Within this picture, the local resistivity maximum at $x_{\mathrm{c(rf)}}$=0.22 may reflect enhanced of carrier scattering corresponding to the AF fluctuations near the boundary at $x \approx$ 0.2 and might contribute to the double-dome-like $T_{\mathrm{c}}$ suppression in the superconducting state of $\lambda$-Ga. This location lies within the low-pressure region where quantum-critical behavior has previously been proposed [14], consistent with these AF fluctuations being the electronic instability associated with that quantum-critical region.

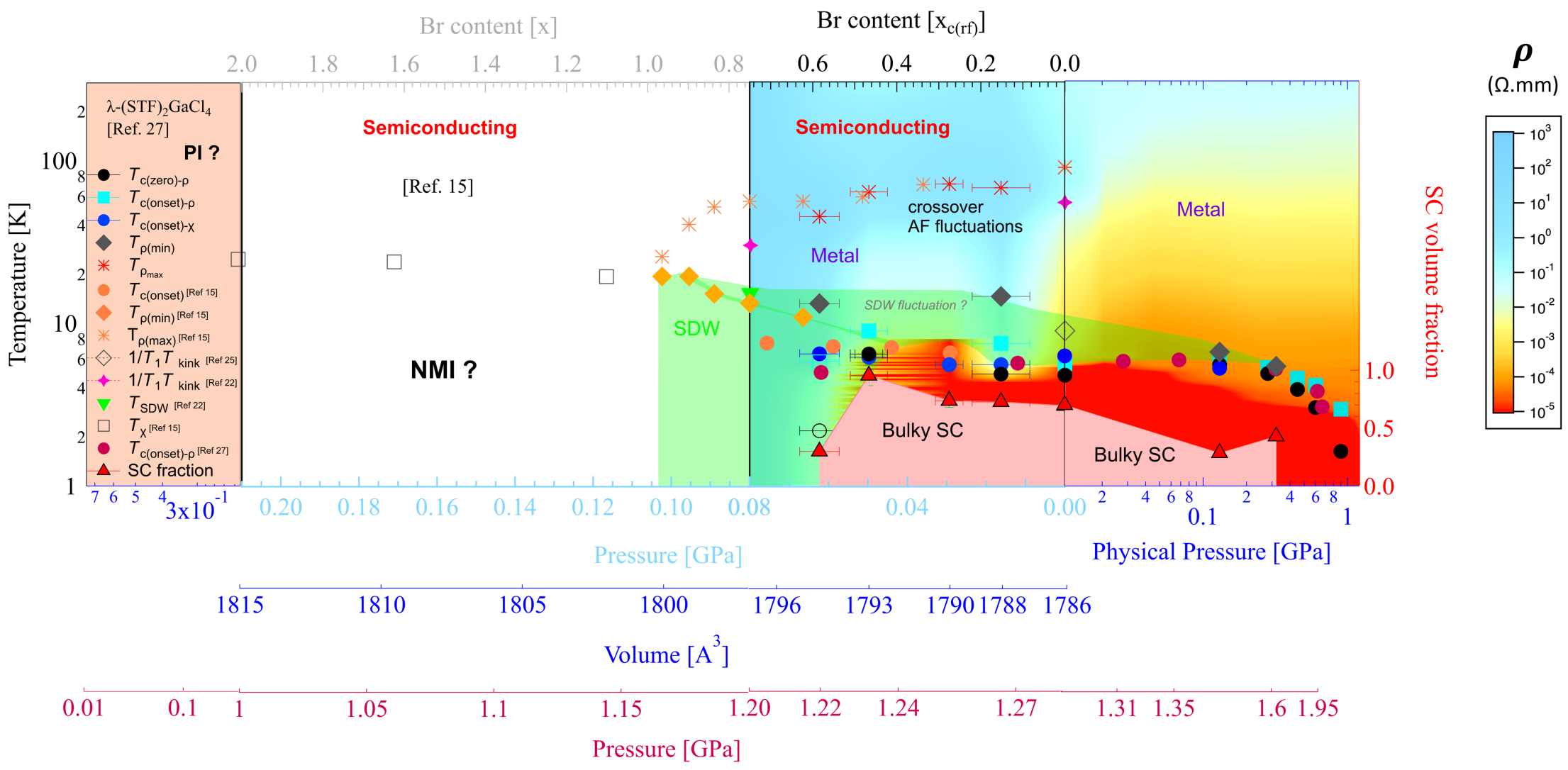

Figure. 10. Pressure–temperature ($P$–$T$) phase diagram of the λ-$(BETS)_2GaCl_4$ family under chemical and physical pressure. Br substitution represents negative chemical pressure through unit-cell expansion. The upper axes show the Br concentration and corresponding unit-cell-volume/pressure scales, while the burgundy axis represents the equivalent pressure scale for λ-$(STF)_2GaCl_4$ [27]. The shaded regions indicate the electronic phases identified from the present and previous studies, including the SDW and bulky SC regions.

Molecular substitution within the λ-$(D)_2GaCl_4$ family ($D$ = ET, STF, BETS) offers a further route to bandwidth control. Replacing ET with STF changes the ground state from an antiferromagnetic insulator to a paramagnetic insulator [16,17], a transition that physical pressure of ~1.65 GPa reproduces in λ-ET [16,17]. In λ-STF, physical pressure above 1.22 GPa induces a superconducting transition [27]. Based on comparable $T_{\mathrm{c}}$ values, λ-STF at 1.35 GPa was previously proposed to correspond to λ-Ga at ambient pressure, under the assumption that superconductivity in λ-Ga disappears above 0.6 GPa [27]. Subsequent resistivity measurements, however, show that superconductivity in λ-Ga persists to 0.9 GPa [13,18], which calls for a reassessment of this correspondence. We find that the resistivity of λ-STF at $P \approx$ 1.22 GPa — where the resistive drop first appears — resembles that of $x_{\mathrm{c(rf)}}$=0.63 [18], so we take 1.22 GPa in λ-STF to correspond to λ-$(BETS)_2GaBr_{0.63}Cl_{3.37}$. On this scale, λ-STF at $P \approx$ 1.27 GPa reproduces the resistivity behavior of $x_{\mathrm{c(rf)}}$=0.22, including its small low-temperature upturn anomaly [27,18]. This agreement suggests that the upturn at $x_{\mathrm{c(rf)}}$=0.22 is intrinsic to the system and supports its identification as the local-resistivity-maximum crossover region described above. The $T_{\mathrm{c(onset)}}$ values of λ-STF (burgundy circles, Fig. 10) overlap with those obtained for λ-Ga under chemical and physical pressure, indicating that chemical substitution, external pressure, and molecular substitution can be integrated into a single bandwidth-controlled phase diagram for λ-type molecular superconductors.

This correspondence is presently based on similarities in $T_c$ and resistivity behavior alone. A direct determination of $U/W$ for the mixed-crystal series is needed to place the unified phase diagram on a quantitative footing, and microscopic probes of the superconducting state at low Br concentration will be essential for establishing its underlying origin.

## 4. Conclusion

We investigated the superconducting magnetic properties of λ-$(BETS)_2GaBr_xCl_{4-x}$ under chemical pressure by magnetic-susceptibility measurements. Br concentrations from EPMA agreed well with single-crystal X-ray diffraction refinement, giving refined compositions $x_{c(rf)}$ = 0.22, 0.24, 0.42, and 0.63, used throughout to discuss the chemical-pressure evolution of superconductivity. All samples exhibited a diamagnetic superconducting transition. The suppressed superconducting volume fraction at $x_{c(rf)}$=0.63 is consistent with proximity to the commensurate spin-density-wave (SDW) state suggested at $x$=0.75 [22] and with the pronounced low-temperature resistivity upturn at $x_{c(rf)}$=0.63 [18]. Notably, the same upturn anomaly recurs in λ-Ga and λ-STF under physical pressure, suggesting that the associated SDW fluctuations may extend continuously from the chemical- to the physical-pressure axis rather than being confined to low Br concentration. From $H_{c1}$ (Meissner-deviation, demagnetization-corrected) and $H_{c2}$ (resistive midpoint, up to 11 T), we constructed the magnetic-field–pressure ($H$–$P$) phase diagram: $H_{c1}$ shows a dome-*like* maximum at $x_{c(rf)}$= 0.24, indicating the strongest resistance to initial magnetic-flux penetration among the investigated compositions, while the available $H_{c2}$ values provide an initial determination of the high-field superconducting boundary on the negative chemical-pressure side. The comparison with the weak-coupling Pauli limit provides a useful reference field scale; however, the present data are insufficient to determine the relative contributions of orbital and paramagnetic pair breaking. Extension of $H_{c2}$ measurements toward the physical-pressure side, together with measurements for both field orientations, will be important for establishing the complete $H$–$P$ phase diagram and clarifying the high-field pair-breaking mechanism.

Combining these magnetic results with previous transport data [13,14,18], we propose an electronic pressure–temperature ($P$–$T$) phase diagram for λ-Ga. Br substitution expands the unit-cell volume and thus acts as negative chemical pressure, in contrast to the κ-$(ET)_2X$ family. The diagram shows a double-dome-like superconducting region with a local minimum at $x$=0, likely arising from competing electronic fluctuations near $x$=0 evidenced by optical-conductivity [24] and $^{13}C$ NMR measurements [25]. This is also consistent with the QCP that has previously been proposed at $P$≈0.13 GPa [14]— separating a low-pressure regime where superconductivity coexists with SDW fluctuations and unconventional Fermi-liquid behavior from a high-pressure regime where superconductivity is followed by conventional Fermi-liquid behavior. This is further supported by the distinct low-Br transport and magnetic behavior, with $x_{c(rf)}$=0.22 as a local resistivity maximum near the boundary between the lower- and higher-resistivity regions in Fig. 10. Combining $T_c$ from λ-$(STF)_2GaCl_4$ under physical pressure with the resistivity correspondence between $P$≈ 1.22 GPa and λ-$(BETS)_2GaBr_{0.63}Cl_{3.37}$, the two-phase diagrams overlap, suggesting that chemical substitution, external pressure, and molecular substitution can be integrated into a unified bandwidth-controlled phase diagram for λ-type molecular superconductors. Direct determination of $U/W$ and microscopic study of the low-Br region remain needed to establish this origin quantitatively.